\documentclass[10pt,twocolumn,showpacs,preprintnumbers,amsmath,amssymb,aps,prl,superscriptaddress,longbibliography]{revtex4-2}
\usepackage{appendix}
\usepackage{bbm}
\usepackage{mathrsfs}
\usepackage{graphicx}
\usepackage{dcolumn}
\usepackage{bm}
\usepackage{braket}
\usepackage{amsmath}
\usepackage{amsfonts}
\usepackage[dvipsnames]{xcolor}
\usepackage[colorlinks=true,linkcolor=Blue,urlcolor=BlueViolet,citecolor=BlueViolet]{hyperref}
\usepackage{natbib}
\usepackage{multirow}
\usepackage{floatrow}
\usepackage{hyperref}

\begin{document}
\title{Symmetry-enabled tunable square-lattice Hubbard models in $\Gamma$-valley moir\'e bilayers}
\author{Rui Shi}
\affiliation{State Key Laboratory of Surface Physics and Department of Physics, Fudan University, Shanghai 200433, China}
\affiliation{Shanghai Research Center for Quantum Sciences, Shanghai 201315, China}
\author{Kejie Bao}
\affiliation{State Key Laboratory of Surface Physics and Department of Physics, Fudan University, Shanghai 200433, China}
\affiliation{Shanghai Research Center for Quantum Sciences, Shanghai 201315, China}
\author{Huan Wang}
\affiliation{State Key Laboratory of Surface Physics and Department of Physics, Fudan University, Shanghai 200433, China}
\affiliation{Shanghai Research Center for Quantum Sciences, Shanghai 201315, China}
\author{Jing Wang}
\thanks{wjingphys@fudan.edu.cn}
\affiliation{State Key Laboratory of Surface Physics and Department of Physics, Fudan University, Shanghai 200433, China}
\affiliation{Shanghai Research Center for Quantum Sciences, Shanghai 201315, China}
\affiliation{Institute for Nanoelectronic Devices and Quantum Computing, Fudan University, Shanghai 200433, China}
\affiliation{Hefei National Laboratory, Hefei 230088, China}
	
\begin{abstract}
Square-lattice moir\'e systems provide a promising route for quantum simulation of Hubbard physics, yet electrically tunable realizations of the hopping ratio $t'/t$ have so far been limited to specific valley configurations. A natural question is whether the same tunability can be extended to the more broadly occurring $\Gamma$-valley systems. Here we identify a symmetry-controlled route for realizing electrically tunable square-lattice Hubbard models in $\Gamma$-valley twisted homobilayers. At small twist angles, an emergent layer-exchange symmetry separates the low-energy states into flat bands localized on two nested square sublattices, suppressing inter-sublattice hopping. An interlayer displacement field breaks this symmetry and induces controllable hybridization between the sublattices, enabling continuous tuning of the $t'/t$ ratio over a wide range while preserving an effective single-band description. We further establish a formal correspondence between $\Gamma$- and M-valley moir\'e systems, revealing them as different symmetry limits of a unified framework for tunable square-lattice Hubbard models. Our results uncover a general symmetry principle underlying displacement-field tunability and extend electrically controllable square-lattice Hubbard physics to a broad family of $\Gamma$-valley materials.
\end{abstract}

\maketitle

Artificial moir\'e superlattices have emerged as a powerful platform for exploring strongly correlated and topological quantum states due to their exceptional \emph{in situ} tunability~\cite{Bistritzer2011,Andrei2021,Kennes2021,balents2020,carr2020,Mak2022,Wu2018}. To date, exotic phenomena including fractional Chern insulators~\cite{Cai2023,Zeng2023,Park2023,Xu2023,Lu2024a,Li2021,Devakul2021,MoralesDuran2023,Wang2024,Jia2024,Yu2024,MoralesDuran2024} and unconventional superconductivity~\cite{Cao2018a,Yankowitz2019,Lu2019,Stepanov2020,Park2021,Hao2021,Xia2024,Guo2025,Qin2025,Tuo2025} have been realized predominantly in hexagonal moir\'e systems, which provide access to many-body physics on triangular and honeycomb lattices~\cite{Mak2026}. More recently, increasing efforts have focused on engineering square-lattice moir\'e systems~\cite{Kariyado2019,Luo2021,Can2021,Soeda2022,Li2022,Song2022,Zhao2023,Volkov2023,Eugenio2023,Xu2025,Sarkar2025,Eugenio2025,Kariyado2025,Bao2026,Liu2025}. The square-lattice Hubbard model is of particular interest because of its central role in correlated-electron physics and its connection to the cuprate high-temperature superconductors~\cite{Scalapino2012,Arovas2022,Qin2022}. Although neutral-atom quantum simulators have realized Hubbard physics in the cryogenic regime~\cite{xu2025neutral}, accessing the low-energy scales and continuously tunable parameter regimes relevant to correlated phases remains challenging. Moir\'e systems, with electrically controllable electronic energy scales, therefore provide a promising route toward tunable quantum simulation of square-lattice Hubbard physics.

A central challenge in realizing such quantum simulators is achieving direct and broad control over Hubbard-model parameters, particularly the ratio of next-nearest- to nearest-neighbor hopping, $t'/t$, which strongly influences the competing correlated phases. Recently, a square-lattice Hubbard model with widely tunable $t'/t$ was proposed in M-valley twisted square homobilayers, where an interlayer displacement field breaks an emergent layer-exchange symmetry and induces controllable inter-sublattice hybridization~\cite{Eugenio2025}. However, experimentally accessible M-valley square-lattice materials remain limited, restricting the material landscape for this tunable Hubbard platform. In contrast, square-lattice materials with band extrema at the $\Gamma$ valley are considerably more abundant and provide a broader opportunity for moir\'e engineering~\cite{Jiang2024a}. This raises a fundamental question: is the symmetry mechanism responsible for displacement-field tunability a special feature of M-valley systems, or can it be generalized to $\Gamma$-valley moir\'e platforms? Thus, identifying the symmetry principle that controls this tunability is essential for expanding moir\'e quantum simulation beyond specific valley configurations.

Here we demonstrate that $\Gamma$-valley twisted square homobilayers realize a general symmetry-based platform for electrically tunable $t$–$t'$–$U$ Hubbard physics. At small twist angles, an emergent layer-exchange symmetry separates the low-energy states into flat bands localized on two nested square sublattices, suppressing inter-sublattice hopping. An interlayer displacement field breaks this symmetry and induces controlled hybridization between the sublattices, providing a direct and continuous tuning knob for the ratio $t'/t$ over a wide range while preserving an effective single-band description. We further establish a formal correspondence between the continuum descriptions of $\Gamma$- and M-valley moir\'e systems, revealing that the M-valley construction represents a high-symmetry limit of a more general framework. This unified perspective identifies the symmetry origin of displacement-field tunability and extends electrically controllable square-lattice Hubbard simulation to a broad family of $\Gamma$-valley materials.

\begin{figure}[b]
	\begin{center}
		\includegraphics[width=\columnwidth, clip=true]{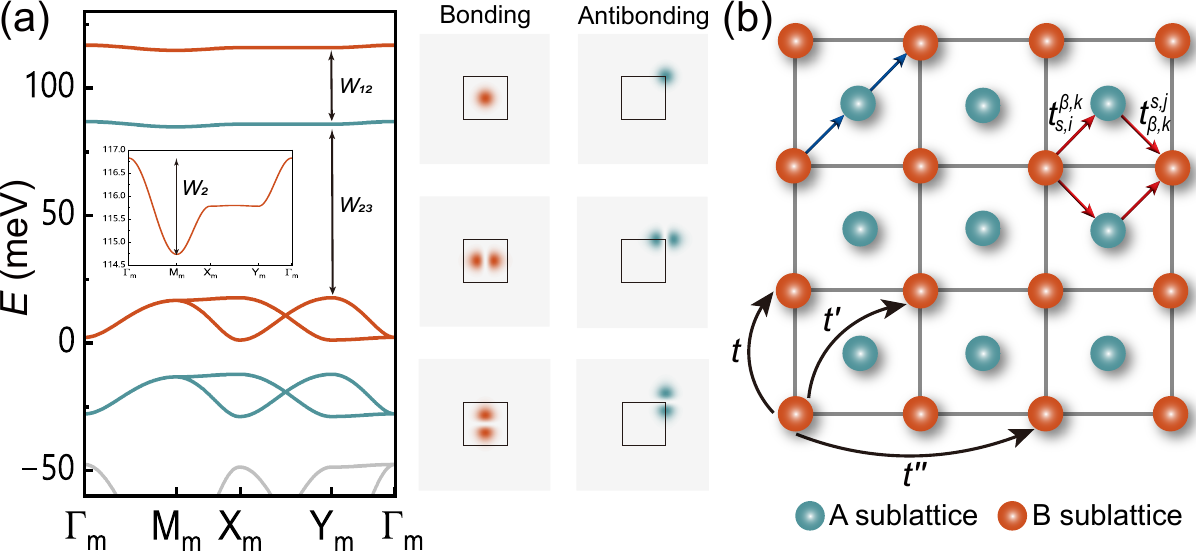}
	\end{center}
	\caption{(a) Band structure of the continuum model at $\vartheta=5^\circ$ with parameters $m=0.6m_{\text{e}}$, $a_0=5$~\AA, $w_0=15$~meV, $w_1=55$~meV, $D=0$. The inset highlights the top flat bands. The two topmost bands share identical dispersion and are split by $2w_0$. The middle panels show the symmetry-adapted maximally localized Wannier functions for the topmost six bands. The black square indicates the moir\'e unit cell. Brown denotes layer-bonding states at the Wyckoff position $(0,0)$, while jasper denotes antibonding states at $(1/2,1/2)$, consistent with the color scheme in (b). (b) Schematic of the moir\'e square superlattice. Layer-bonding and antibonding states reside at B and A sublattice, respectively. The parameters $t$, $t'$, and $t''$ denote the first-, second-, and third-nearest-neighbor hoppings of the effective single-band model. Red and blue arrows indicate second-order virtual processes contributing to $t$ and $t'$, respectively.}
	\label{fig1}
\end{figure}

\emph{Model---}We consider twisted square-lattice homobilayers at small twist angle $\vartheta$ using a continuum description focused on the single-band manifold near the $\Gamma$ point of the Brillouin zone (BZ). This is different from the twisted FeSe bilayers, which require multiband treatment at the same high-symmetry point~\cite{Eugenio2023}. The moir\'e lattice vectors are $\mathbf{L}_i = \hat{\mathbf{z}}\times \mathbf{a}_i/[2\sin(\vartheta/2)]$, with the moir\'e period $\ell_m=a_0/[2\sin(\vartheta/2)]$, where $a_0$ is the monolayer lattice constant. The continuum Hamiltonian takes the form
\begin{equation}\label{eq1}
	\mathcal{H}_0 = \begin{pmatrix}
		\frac{\boldsymbol{\nabla}^2}{2m} & \Delta_T(\mathbf{r})\\
		\Delta_T^\dagger(\mathbf{r}) & \frac{\boldsymbol{\nabla}^2}{2m}
	\end{pmatrix},
\end{equation}
where $m$ is the effective mass. $\Delta_T(\mathbf{r})$ describes the moir\'e-periodic interlayer tunneling. Under the minimal $D_4$ symmetry and time-reversal symmetry $\mathcal{T}$, the leading harmonic contribution is
\begin{equation}
\Delta_T(\mathbf{r})\equiv w_0+t(\mathbf{r})=w_0 + 2w_1\text{cos}(\mathbf{g}_1\cdot \mathbf{r}) + 2w_1\text{cos}(\mathbf{g}_2\cdot \mathbf{r}),
\nonumber
\end{equation}
where $w_0$ and $w_1$ are real parameters denoting the uniform and spatially modulated tunneling amplitudes, respectively, and $\mathbf{g}_i$ are moir\'e reciprocal lattice vectors satisfying $\mathbf{g}_i\cdot\mathbf{L}_j=2\pi\delta_{ij}$. Spin-dependent tunneling is neglected because the combined $C_{2z}\mathcal{T}$ symmetry enforces an effective spin $SU(2)$ invariance.

We focus on the regime $w_0<w_1$, where an emergent layer-exchange symmetry gives rise to a pair of isolated flat bands forming the basis of a tunable single-band square-lattice model. Two characteristic quantities control this regime. First, below the characteristic twist angle $\vartheta^{*}=\sqrt{2mw/(\sqrt{2}\pi/a_0)^2}$ with $w=\text{min}\{4w_1 + w_0, 4w_1-w_0\}$, two isolated flat bands emerge. Second, the uniform tunneling $w_0$  determines their energy splitting and exceeds the bandwidth $W_2$ in the parameter regime considered. For the representative parameters in Fig.~\ref{fig1}(a), $\vartheta/\vartheta^*=0.43$, the two highest bands are nearly flat, well separated from other bands ($W_{23}>W_2$), and exhibit identical dispersions. Projection onto either band gives the square-lattice tight-binding dispersion,\begin{equation}\label{eq2}
\begin{aligned}
\epsilon_{\mathbf{k}} &= 2t\left(\cos (k_x) + \cos(k_y) \right) \\
        &\quad+ 2t'\left(\cos(k_x+k_y) + \cos(k_x-k_y)\right),
\end{aligned}
\end{equation}
with $t=0.26$~meV and $t'= 0$.

The origin of these two decoupled flat bands is an emergent layer-exchange symmetry $\tau_x$, satisfying $[\mathcal{H}_0,\tau_x]=0$. The eigenstates can therefore be classified as layer-bonding and antibonding combinations, $(1,\pm1)^T\psi_\pm(\mathbf{r})$, where $\psi_\pm$ obey
\begin{equation}\label{eq3}
\left[\frac{\boldsymbol{\nabla}^2}{2m} \pm \Delta_T(\mathbf{r})\right]\psi_\pm(\mathbf{r})
= (E \pm w_0)\psi_\pm(\mathbf{r}).
\end{equation}
Because $t(\mathbf{r}+(\mathbf{L}_1\pm\mathbf{L}_2)/2)=-t(\mathbf{r})$, the bonding and antibonding wave functions are related by a half translation: $\psi_-(\mathbf{r})=\psi_+(\mathbf{r}+(\mathbf{L}_1\pm\mathbf{L}_2)/2)$, and their energy shift is $2w_0$, with band gap between them defined as $W_{12}$. In the small-angle limit, the kinetic energy scale decreases as $\sim 1/|\mathbf{L}_i|^2$, while the moir\'e tunneling dominates. Consequently, the low-energy physics is governed by $|\Delta_T(\mathbf{r})|$, whose maxima trap the holes. When $w_1>w_0$, the maxima occur at the two inequivalent Wyckoff positions $(0,0)$ and $(1/2,1/2)$, corresponding respectively to the bonding and antibonding states. The resulting Wannier orbitals form two nested square sublattices (A and B in Fig.~\ref{fig1}(b)), and inter-sublattice hopping is forbidden by the emergent layer-exchange symmetry.

The intralayer moir\'e potential, $V(\mathbf{r})\tau_0$, is also allowed by symmetry. A layer-symmetric intralayer potential preserves the decoupling of the two parity sectors, although it generally modifies their dispersions and energy separation. For $V(\mathbf r)\ll w_1$, it only provides quantitative corrections to the band parameters; therefore we set $V(\mathbf{r})=0$ in the following.

To confirm the symmetry-derived picture, we construct symmetry-adapted maximally localized Wannier functions (MLWFs)~\cite{Marzari1997,Marzari2012,Sakuma2013,Pizzi2020} for the highest six bands shown in Fig.~\ref{fig1}(a). The resulting MLWFs reveal that the bonding states contain $s$, $p_x$, $p_y$ orbitals centered at the Wyckoff position $(0,0)$, whereas the antibonding states have the same orbital character centered at $(1/2,1/2)$, directly confirming the emergent nested square-sublattice structure.

\begin{figure}[b]
	\begin{center}
		\includegraphics[width=\columnwidth, clip=true]{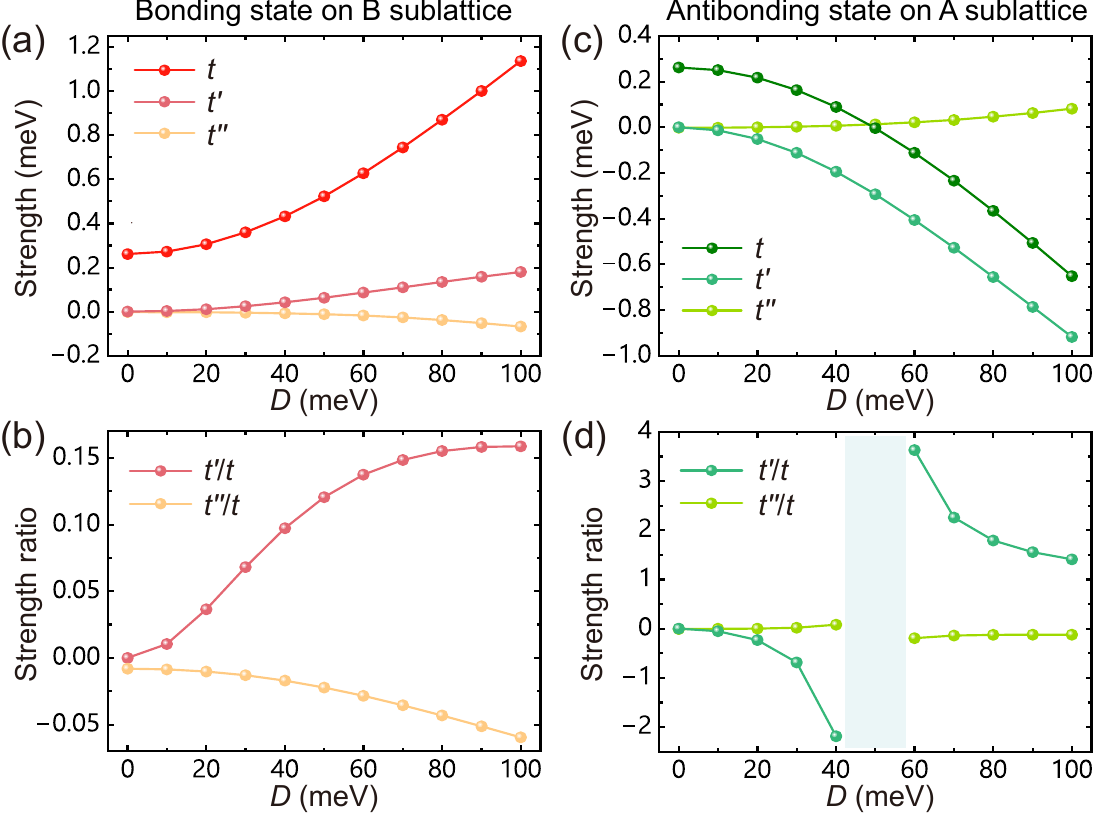}
	\end{center}
	\caption{Dependence of the hopping parameters $t$, $t'$, $t''$ and their ratios $t'/t$, $t''/t$ on the displacement field $D$. (a),(b) Results for the topmost band (layer-bonding state localized on the B sublattice). (c),(d) Results for the second topmost band (layer-antibonding state localized on the A sublattice). The gray shaded region in (d) indicates the regime where $t\approx 0$, leading to a divergence of $t'/t$. The continuum model parameters are the same as in Fig.~\ref{fig1}.}
	\label{fig2}
\end{figure}

\emph{Tunable $t$-$t'$ model---}The layer-exchange symmetry responsible for the decoupling of the two moir\'e sublattices also provides a direct route toward engineering longer-range hopping. At zero displacement field, inter-sublattice hybridization is forbidden by this symmetry, resulting in vanishing nearest-neighbor hopping between the two emergent sublattices. Applying a displacement field $D\tau_z$ breaks the layer-exchange symmetry and activates controlled hybridization between bonding and antibonding states. When $W_{23},W_{12}>W_2$, the low-energy physics can be projected onto either the topmost bonding or antibonding band, yielding an effective single-band square-lattice model with field-dependent hopping parameters.

The displacement field preserves $C_{4z}$ symmetry while breaking $C_{2x}$ and $C_{2y}$, and the emergent layer-exchange symmetry. Within the regime where the bands remain isolated, we construct symmetry-adapted MLWFs~\cite{supple} and extract the effective hoppings $t$, $t'$, and $t''$. Fig.~\ref{fig2}(a,b) show their evolution for the layer-bonding state. The ratio $t'/t$ continuously increases from zero to $0.16$, covering the parameter regime where unconventional correlated phases, including superconducting states, have been predicted in the $t$-$t'$-$U$ Hubbard model~\cite{Gong2021,Jiang2021a}.

The layer-antibonding state exhibits even stronger tunability. As shown in Fig.~\ref{fig2}(c,d), $t'/t$ can be tuned continuously across both negative and positive values, with a divergence near $D\simeq50$~meV where the nearest-neighbor hopping vanishes. Moreover, for $D \gtrsim 32$ meV, $|t'|>|t|$, while the third-nearest-neighbor hopping $t''$ remains strongly suppressed. Thus, the displacement field provides a direct experimental control parameter for continuously tuning the effective hopping parameters of a square-lattice Hubbard model.

The field dependence of the hopping parameters can be understood from a perturbative picture. We construct a six-band tight-binding model by projecting the continuum Hamiltonian onto the subspace consisting of the highest six bands, including the $s,p_x,p_y$ orbitals of both bonding and antibonding states. This description remains valid as long as these bands remain energetically isolated during the variation of $D$. The field-induced correction to the hopping amplitude between $s$ orbitals at sites $i$ and $j$ in sublattice $\ell=\text{A}$ or B is
\begin{equation}\label{eq4}
\Delta t^{\ell}_{ji} 
= \delta t^{\ell}_{ji}(D)
+ \sum_{k\in\ell'}\sum_{\beta=s,p_x,p_y}
\frac{t^{s,j}_{\beta,k}\, t^{\beta,k}_{s,i}}{E^\ell_{s} - E^{\ell'}_{\beta}}.
\end{equation}
The first term describes the direct modification of hopping caused by Wannier-function deformation under the displacement field. The second term arises from second-order virtual processes enabled by the field-induced inter-sublattice hybridization. Here $\ell'$ denotes the neighboring sublattice of $\ell$, while $E^{\ell}_{s}$ and $E^{\ell'}_{\beta}$ are the corresponding onsite energies, $t^{s,j}_{\beta,k}$ and $t^{\beta,k}_{s,i}$ denote inter-sublattice hopping amplitudes enabled by the breaking of layer-exchange symmetry. The relevant virtual hopping paths contributing to $t$ and $t'$ are illustrated in Fig.~\ref{fig1}(b).

\begin{figure}[b]
	\begin{center}
		\includegraphics[width=\columnwidth, clip=true]{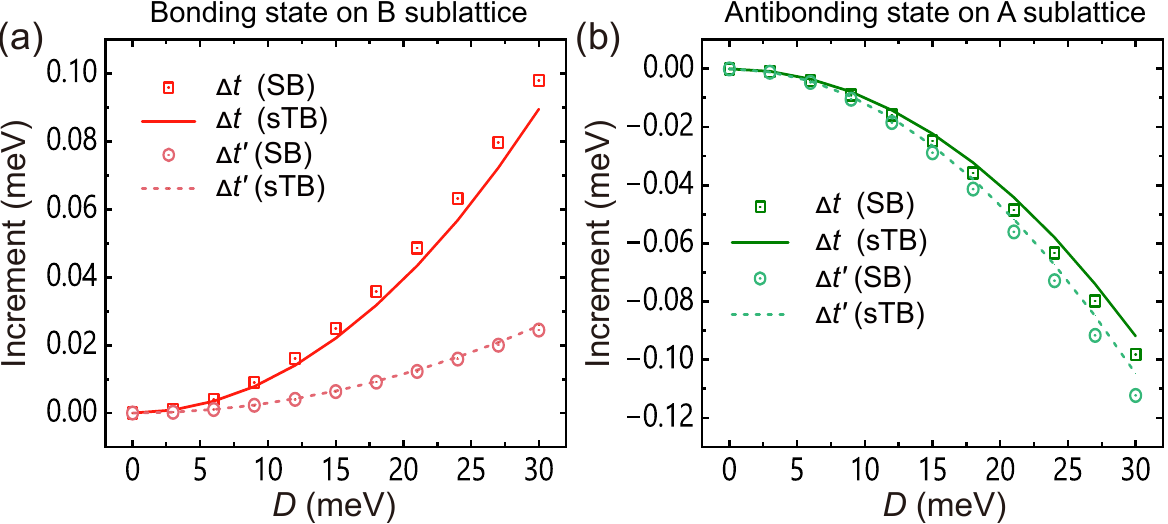}
	\end{center}
	\caption{Perturbative understanding of the displacement-field dependence of the hopping parameters $t$ and $t'$. (a) Layer-bonding state. (b) Layer-antibonding state. Here $\Delta t \equiv t(D)-t(0)$ and $\Delta t' \equiv t'(D)-t'(0)$. Symbols (squares and circles) denote $\Delta t$ and $\Delta t'$ obtained from single-band (SB) projections of the continuum model, as shown in Fig.~\ref{fig2}. Solid and dashed lines represent the corresponding results for $\Delta t$ and $\Delta t'$, respectively, calculated perturbatively from the six-band tight-binding (sTB) model. In the shown range of $D$, the topmost six bands remain well isolated from higher-energy bands.}
	\label{fig3}
\end{figure}

For the layer-antibonding states on sublattice A, both $\Delta t$ and $\Delta t'$ are negative and comparable in magnitude ($\Delta t \sim \Delta t'$). This behavior is dominated by second-order virtual processes, which govern the field dependence of the hopping parameters. Specifically, the combined effects of onsite-energy differences and orbital anisotropy between the $s$ and $p_x,p_y$ orbitals on sublattice B lead to nearly equal negative corrections to both $t$ and $t'$. Given that the initial value of $t$ is positive, the field-induced correction can drive $t$ through zero, thereby enabling a wide range of tunability for the $t'/t$ ratio. In contrast, for the layer-bonding states on sublattice B, both $\Delta t$ and $\Delta t'$ are positive, with $\Delta t > \Delta t'$. This asymmetry arises from the competing contributions of $s$ and $p_x,p_y$ orbitals on sublattice A, together with the additional direct correction $\delta t^{\ell}_{ji}(D)$. Further details are provided in Supplemental Material~\cite{supple}.

Fig.~\ref{fig3} compares the hopping corrections obtained from direct projection of the continuum model and from the perturbative six-band model. The excellent agreement confirms that the displacement-field dependence of $t$ and $t'$ originates from symmetry-breaking-induced virtual hopping processes. The total corrections $\Delta t$ and $\Delta t'$ are both proportional to $D^2$ due to $\tau_{x} \mathcal{H}_0(D)\tau_{x}=\mathcal{H}_0(-D)$, which enforces the effective hoppings to be even functions of \(D\) for an isolated, nondegenerate band~\cite{supple}. Deviations for $D \gtrsim 30$~meV arise from hybridization of the $p_x,p_y$ orbitals with higher-energy bands, where corrections beyond the six-band approximation become important.

\begin{figure}[t]
	\begin{center}
		\includegraphics[width=\columnwidth, clip=true]{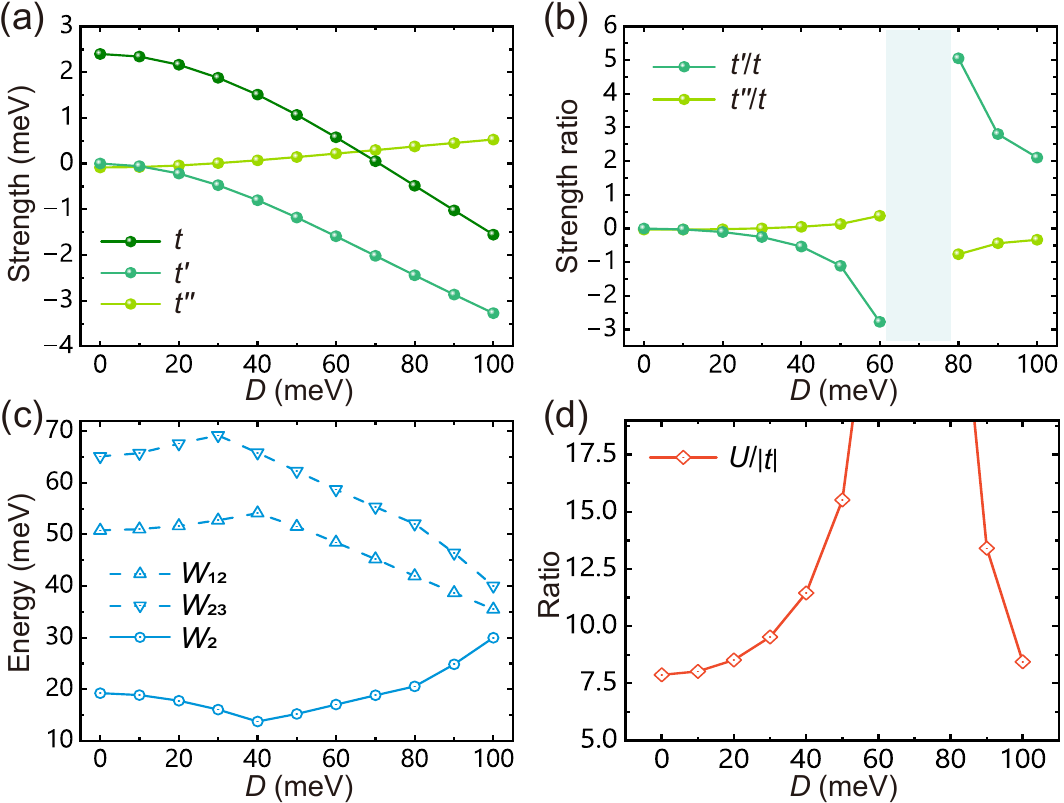}
	\end{center}
	\caption{$t$-$t'$-$U$ model for the second topmost band. (a),(b) Displacement-field dependence of the hopping parameters $t$, $t'$, $t''$ and their ratios $t'/t$, $t''/t$. (c) Bandwidth of the second topmost band ($W_2$) and its energy separations from the first ($W_{12}$) and third ($W_{23}$) bands. (d) Ratio $U/|t|$, obtained by projecting the screened Coulomb interaction onto the second topmost band with $\epsilon=75$ and $\xi=3$~nm. All results are calculated for $\vartheta=8^\circ$, $m=0.6m_{\mathrm{e}}$, $w_0=35$~meV, and $w_1=75$~meV.}
	\label{fig4}
\end{figure}

\emph{Single-band $t$-$t'$-$U$ Hubbard model---}We now demonstrate the realization of a single-band $t-t'-U$ Hubbard model in the second topmost band, which exhibits strong displacement-field tunability. To obtain a sizable hopping scale while maintaining well-isolated low-energy bands, we choose the continuum parameters $\vartheta=8^\circ$, $m=0.6m_e$, $w_0=35$~meV, and $w_1=75$~meV. Fig.~\ref{fig4}(a,b) show the displacement-field evolution of the hopping parameters and their ratios, following the same symmetry-controlled mechanism discussed above. 

To quantify the interaction strength, we project the screened Coulomb interaction onto the Wannier functions of the antibonding state. The resulting interaction matrix elements are
\begin{equation}
	U(\mathbf{R}) = \int d\mathbf{r}d\mathbf{r}' \rho(\mathbf{r}) \tilde{U}(\mathbf{r} + \mathbf{R} - \mathbf{r}')\rho(\mathbf{r}'),
\end{equation}
where $\rho(\mathbf{r})$ is the Wannier density and $\mathbf{R}$ denotes the separation between Wannier orbitals. We consider a dual-gate screened Coulomb interaction~\cite{Bernevig2021}, given by
\begin{equation}
	\tilde{U}(\mathbf{r}) = U_{\xi}\sum_{n=-\infty}^{\infty}\frac{(-1)^n}{\sqrt{(r/\xi)^2 + n^2}},
\end{equation}
with $U_{\xi} = e^2/(4\pi\epsilon_0 \epsilon \xi)$, where $\epsilon$ is the dielectric constant and $\xi$ is the gate distance, and $r = |\mathbf{r}|$. The calculated interaction matrix elements $U(\mathbf{R})$ decay exponentially with increasing $|\mathbf{R}|$ and the onsite interaction $U\equiv U(0)$ dominates over longer-range interactions~\cite{supple}, establishing an effective single-band Hubbard description.

Fig.~\ref{fig4}(b) shows that $t'/t$ can be continuously tuned from $0$ to $-1.1$ as the displacement field increases from $0$ to $50$~meV, covering the commonly used cuprate regime $t'/t\approx-0.3$~\cite{Hirayama2018,Hirayama2019}. The interaction strength is characterized by $U/t$, shown in Fig.~\ref{fig4}(d). Because the nearest-neighbor hopping decreases under the displacement field, $U/t$ increases from approximately $8$ to $15.5$ within this range for $\epsilon\approx 75$. These values are comparable to those estimated for cuprate superconductors~\cite{Hirayama2018,Hirayama2019}. Such dielectric screening can be achieved using high-permittivity substrates such as SrTiO$_3$~\cite{Pai2018}. Throughout this parameter range, the single-band description remains valid: the separation from neighboring bands exceeds the bandwidth by a factor larger than $2.5$, and the interaction strength remains below the band gap.

\emph{Parameters justification---}The emergence of a tunable single-band Hubbard model in $\Gamma$-valley twisted bilayers requires a parameter regime where $w_0$ is moderate and $w_0 < w_1$. This hierarchy is essential to preserve the single-band description while maintaining the exceptional tunability characterized in our analysis. While it has been argued that $w_0$ dominates in $\Gamma$-valley bands~\cite{Eugenio2025}, such a conclusion may not hold in realistic materials where lattice reconstruction is significant upon twisting. Indeed, $w_1$ can exceed $w_0$ in practical systems, as calculated in twisted bilayer black phosphorus~\cite{Wang2023}. To support this, we performed large-scale \emph{ab initio} calculations for twisted bilayer Cu$_2M$Se$_4$ ($M$ = Mo, W), where the monolayer crystallizes in a tetragonal lattice with a conduction band minimum at $\Gamma$. As detailed in Supplemental Material~\cite{supple}, we extract moir\'e potential parameters $w_0=4$~meV, $w_1=40$~meV for Cu$_2$MoSe$_4$ and $w_0=8$~meV, $w_1=50$~meV for Cu$_2$WSe$_4$, confirming that the parameter regime required for tunable Hubbard physics can naturally occur in realistic $\Gamma$-valley materials.

\emph{Discussion---}The two topmost bands, derived from the $s$-orbitals of the layer-bonding and antibonding states, provide an effective single-band $t$–$t'$–$U$ Hubbard model with electrically tunable parameters. The central feature of this platform is the wide-range control of the next-nearest-neighbor to nearest-neighbor hopping ratio $t'/t$, enabled by displacement-field-induced hybridization between the two nested square sublattices. This provides a direct experimental knob for exploring how longer-range hopping reshapes square-lattice Hubbard physics. In particular, numerical studies have shown that $t'$ strongly influences the competition between $d$-wave superconductivity and stripe orders~\cite{Ponsioen2019,Jiang2021a,Jiang2024,Lu2024b,Xu2024}. The present platform therefore enables systematic studies of these theoretically distinct regimes within a single material system, without requiring a change of lattice geometry or chemical composition.

The accessible parameter space also overlaps with regimes associated with other correlated phases. For example, the range $0.68\leq |t'/t|\leq0.72$ (or equivalently $0.46 \leq J'/J \leq 0.52$ in $J$-$J'$ model) has been proposed to support quantum spin liquid behavior~\cite{Jiang2012,Nomura2021,Hu2013,Gong2014,Wang2016,Wang2018,Jiang2021}. The large moir\'e length further provides access to appreciable magnetic flux per moir\'e unit cell under experimentally accessible magnetic fields, providing opportunities to investigate the interplay between topology and strong correlations in square lattices~\cite{Affleck1988,Wen1989,Hatsugai1990}. The interaction strength provides an additional tuning parameter: while our representative calculations yield $U/t\sim10$ for $\epsilon=75$, weaker dielectric screening can increase $U/t$ and access stronger interaction regimes, including antiferromagnetic and other interaction-driven phases~\cite{Nagaoka1966,Arovas2022}. Together, these results demonstrate electrical control over a broad region of square-lattice Hubbard parameter space within a single moir\'e platform.

The relevant energy scales are also favorable for low-temperature quantum simulation. For the $t$–$t'$–$J$ model, numerical calculations estimate $d$-wave superconductivity to persist up to $T_c\sim0.05t$ near optimal doping~\cite{Qu2024}. In our system, $t$ decreases from approximately $2.4$ to $1$~meV as the displacement field is increased to $D=50$~meV [Fig.~\ref{fig4}(a)], corresponding to an energy scale of order several hundred millikelvin for such a transition. This estimate highlights the relatively large energy scales of moir\'e Hubbard platforms compared with cryogenic quantum simulators. The actual transition temperature will depend on microscopic effects beyond the effective Hubbard description, including electron-phonon coupling, and requires further theoretical and experimental investigation.

Finally, we establish a unified framework connecting $\Gamma$-valley and M-valley moir\'e systems~\cite{Eugenio2025,Kariyado2025}, with the details provided in End Matter. By doubling the monolayer unit cell and folding the BZ corner to the center, the M-valley continuum model can be mapped onto the $\Gamma$-valley Hamiltonian under specific constraints. In particular, the M-valley limit corresponds to vanishing zeroth-order interlayer tunneling, $w_0=0$, which preserves the degeneracy between layer-bonding and antibonding states. The M-valley construction is therefore a symmetry-constrained subclass of the more general $\Gamma$-valley formulation. For finite $w_0$, the $\Gamma$-valley system lifts this degeneracy and allows the field-induced corrections to the effective hopping parameters to be described by non-degenerate perturbation theory, whereas the M-valley limit requires degenerate perturbation theory. This correspondence identifies the common symmetry origin of displacement-field tunability in the two valley configurations while extending the experimentally relevant materials space to stable and exfoliable $\Gamma$-valley square-lattice materials~\cite{Jiang2024a}. More broadly, it establishes $\Gamma$-valley twisted bilayers as a general platform for electrically tunable square-lattice Hubbard physics.

\begin{acknowledgments}
\emph{Acknowledgments---}We thank Yang Qi and Yuanbo Zhang for valuable discussions. This work is supported by the Natural Science Foundation of China through Grant No.~12350404, National Key Research Program of China under Grant No.~2025YFA1411400, Quantum Science and Technology-National Science and Technology Major Project through Grant No.~2021ZD0302600, the Science and Technology Commission of Shanghai Municipality under Grants No.~23JC1400600, No.~24LZ1400100 and No.~2019SHZDZX01, and it is sponsored by the ``Shuguang Program'' supported by the Shanghai Education Development Foundation and Shanghai Municipal Education Commission. 

R.S. and K.B. contributed equally to the work.
\end{acknowledgments}

\emph{Note added.} A related study~\cite{Kariyado2026}
develops single- and two-orbital Hubbard descriptions of
$\Gamma$-valley square-lattice moir\'e systems. The present work examines displacement-field-controlled hopping in isolated bonding- and antibonding-derived bands and establishes a symmetry-constrained correspondence with the M-valley construction.

\onecolumngrid
\clearpage
\twocolumngrid

\appendix*
\setcounter{equation}{0}

\begin{center}
{\bf\large End Matter}
\end{center}

\subsection{Connection between $\Gamma$-valley and M-valley moir\'e}

This section provides a unified framework for the continuum Hamiltonians of $\Gamma$-valley and M-valley systems, elucidating the origin of Hubbard model tunability in both platforms. We demonstrate that M-valley twisted bilayers can be mapped onto the $\Gamma$-valley description through a formal unit-cell doubling and BZ folding procedure. This mapping reveals that the M-valley continuum model is a high-symmetry realization of the $\Gamma$-valley formalism, subject to specific emergent constraints.

\begin{figure}[b]
	\begin{center}
		\includegraphics[width=\columnwidth, clip=true]{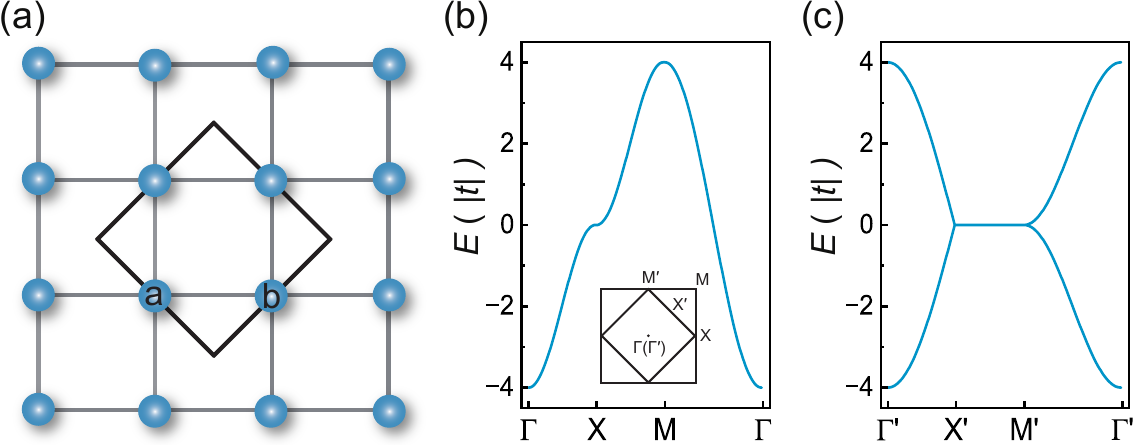}
	\end{center}
	\caption{Lattice and band structures of monolayer. (a) Schematic of the 2D tight-binding model. The gray and black squares mark the primitive and enlarged unit cells, respectively. (b),(c) Band structures corresponding to the primitive and enlarged unit cells, respectively. The inset in (b) shows the original and reduced Brillouin zone.}
	\label{fig1_EM}
\end{figure}

As a starting point, we consider a square-lattice monolayer with a single $s$-orbital per primitive unit cell [gray square, Fig.~\ref{fig1_EM}(a)]. With nearest-neighbor hopping  $t < 0$, the band extremum is located at the M points of the BZ [Fig.~\ref{fig1_EM}(b)]. Alternatively, the system can be described using a $\sqrt{2}\times\sqrt{2}$ enlarged unit cell [black square, Fig.~\ref{fig1_EM}(a)] containing two sublattices ($a$ and $b$). In this representation, the M point folds to the $\Gamma$ point of the reduced BZ, and the resulting band structure is obtained by folding the original bands into the reduced BZ, as illustrated in Fig.~\ref{fig1_EM}(c). Crucially, the highest band edge at the folded $\Gamma$-point now consists of the antibonding combination of the $a$ and $b$ sublattices—a feature that remains robust even in the presence of long-range hopping.

We now compare two perspectives for the twisted bilayer system at angle $\vartheta$, which naturally establishes the connection between $\Gamma$- and M-valley moir\'e description.

\begin{figure*}[t]
	\begin{center}
		\includegraphics[width=0.85\columnwidth, clip=true]{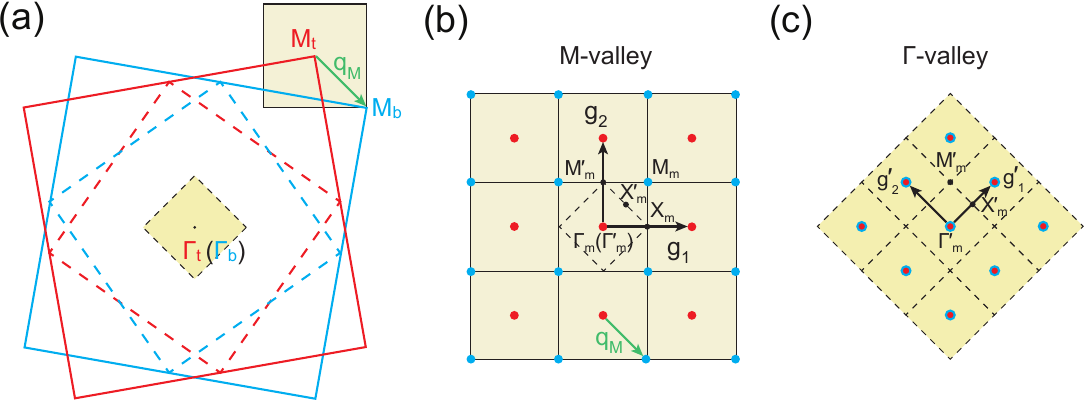}
	\end{center}
	\caption{The moir\'e Brillouin zone. (a) The solid (dashed) red and blue squares represent the BZ of the top and bottom layers for the original (enlarged) lattice. The red and blue dots in (b) and (c) mark the M and $\Gamma$ points of the two layers, forming the mBZ for the M-valley and $\Gamma$-valley, respectively.}
	\label{fig2_EM}
\end{figure*}

\subsubsection{Perspective \uppercase\expandafter{\romannumeral1}: M-valley}
In this perspective, we construct the M-valley moir\'e Hamiltonian directly from symmetry considerations. As illustrated in Figs.~\ref{fig2_EM}(a) and Fig.~\ref{fig2_EM}(b), the low-energy states near the M points of the two layers constitute the M-valley moir\'e Brillouin zone (mBZ). The moir\'e reciprocal lattice vectors $\mathbf{g}_i$ satisfy $\mathbf{L}_{i} \cdot \mathbf{g}_j = 2\pi\delta_{ij}$, where the moir\'e lattice vectors are $\mathbf{L}_{i} = \hat{z} \times \mathbf{a}_i / [2\sin(\vartheta/2)]$.  In the continuum real-space basis $c^\dagger_{\text{M},l,\mathbf{r}}$ (where $l = t, b$ denotes layer index), the single-particle Hamiltonian is given by
\begin{equation}\label{EMeq1}
\begin{aligned}
    \hat{\mathcal{H}}_{\text{M}} = & \int d^2r\ (c^\dagger_{\text{M},t,\mathbf{r}},c^\dagger_{\text{M},b,\mathbf{r}})\mathcal{H}_0^{\text{M}}\begin{pmatrix}
        c_{\text{M},t,\mathbf{r}}\\
        c_{\text{M},b,\mathbf{r}}
    \end{pmatrix},\\
    \mathcal{H}_0^{\text{M}} = &\begin{pmatrix}\frac{\boldsymbol{\nabla}^2}{2m}+ V_{\text{M},t}(\mathbf{r})& \Delta_{\text{M},T}(\mathbf{r})\\
    \Delta_{\text{M},T}^\dagger(\mathbf{r})& \frac{\boldsymbol{\nabla}^2}{2m}+ V_{\text{M},b}(\mathbf{r})\end{pmatrix}.
\end{aligned}
\end{equation}

The translation operator $T_{\mathbf{R}}$ acts on the basis as
\begin{equation}\label{EMeq2}    
T_{\mathbf{R}}c^\dagger_{\text{M},l,\mathbf{r}}T_{\mathbf{R}}^{-1} =e^{-i\Delta \mathbf{M}_l\cdot\mathbf{R}}c^\dagger_{\text{M},l,\mathbf{r+R}},
\end{equation}
where $\mathbf{R} \in \{m\mathbf{L}_{1}+n\mathbf{L}_{2}|m,n\in \mathbb{Z}\}$ is the moir\'e lattice vector, $\Delta \mathbf{M}_t=0$ and $\Delta \mathbf{M}_b=\frac{1}{2}(\mathbf{g}_{1}-\mathbf{g}_{2})\equiv\mathbf{q}_{\text{M}}$. This layer-dependent phase $e^{-i\Delta \mathbf{M}_l\cdot \mathbf{R}}$ arises from the relative shift of the M points between layers. With the moir\'e translational symmetry, the potentials $V_{\text{M},l}(\mathbf{r})$ and the modified tunneling term $e^{-i\mathbf{q}_{\text{M}}\cdot\mathbf{r}}\Delta_{\text{M},T}(\mathbf{r})$ are periodic with respect to the moir\'e superlattice and are expanded in the moir\'e reciprocal lattice vectors $\mathbf{G} \in \{m\mathbf{g}_{1} + n\mathbf{g}_{2}|m,n\in\mathbb{Z}\}$ as 
\begin{equation}\label{EMeq3}
    \begin{aligned}
        V_{\text{M},l}(\mathbf{r}) =& \sum_{\mathbf{G}}e^{i\mathbf{G}\cdot\mathbf{r}}V_{\text{M},l}(\mathbf{G}),\\
        \Delta_{\text{M},T}(\mathbf{r})=&\sum_{\mathbf{G}}e^{i(\mathbf{q}_{\text{M}}+\mathbf{G})\cdot\mathbf{r}}\Delta_{\text{M}}(\mathbf{q}_{\text{M}} + \mathbf{G}).
    \end{aligned}
\end{equation}
Other crystalline symmetries will further constrain the expansion coefficients $V_{\text{M}}(\mathbf{G})$ and $\Delta_{\text{M}}(\mathbf{q}_\text{M} + \mathbf{G})$.

\subsubsection{Perspective \uppercase\expandafter{\romannumeral2}: $\Gamma$-valley}
Alternatively, we fold the monolayer band maximum from the M point to the $\Gamma$ point by doubling the monolayer unit cell (see Fig.~\ref{fig1_EM}(c)), then the moir\'e Hamiltonian for twisted bilayer can be constructed based on $\Gamma$-valley.  The resulting $\Gamma$-valley moir\'e lattice vectors $\mathbf{L}'_{i}$ and $\mathbf{g}'_{i}$, satisfying $\mathbf{L}'_{i}\cdot\mathbf{g}'_{j} = \delta_{ij}$, are related to the M-valley parameters by $\mathbf{L}'_{1,2} = \mathbf{L}_{1}\pm\mathbf{L}_{2}$. The single-particle Hamiltonian is given by
\begin{equation}\label{EMeq4}
\begin{aligned}
    \hat{\mathcal{H}}_\Gamma = & \int d^2r\ (c^\dagger_{\Gamma,t,\mathbf{r}},c^\dagger_{\Gamma,b,\mathbf{r}})\mathcal{H}_0^\Gamma\begin{pmatrix}
        c_{\Gamma,t,\mathbf{r}}\\
        c_{\Gamma,b,\mathbf{r}}
    \end{pmatrix},\\
    \mathcal{H}^\Gamma_0 = & \begin{pmatrix}\frac{\boldsymbol{\nabla}^2}{2m}+ V_{\Gamma,t}(\mathbf{r})& \Delta_{\Gamma,T}(\mathbf{r})\\
    \Delta_{\Gamma,T}^\dagger(\mathbf{r})& \frac{\boldsymbol{\nabla}^2}{2m}+ V_{\Gamma,b}(\mathbf{r})\end{pmatrix}.
\end{aligned}
\end{equation}
The translation operator $T_{\mathbf{R}'}$ acts on the basis as
\begin{equation}\label{EMeq5}            
T_{\mathbf{R'}}c^\dagger_{\Gamma,l,\mathbf{r}}T_{\mathbf{R'}}^{-1} =c^\dagger_{\Gamma,l,\mathbf{r+R'}},
\end{equation}
with $\mathbf{R'} \in \{ m\mathbf{L}'_{1} + n\mathbf{L}'_{2} \mid m,n \in \mathbb{Z} \}$. Since unit-cell enlargement is a formal construction, the $\Gamma$-valley description possesses emergent fractional translation symmetries. Specifically, the system is invariant under a fractional translation symmetry $T_{\mathbf{R}_h}$ (where $\mathbf{R}_h \in \mathbf{R}$ and $\mathbf{R}_h \notin \mathbf{R'}$), acting as
\begin{equation}\label{EMeq6}            
T_{\mathbf{R}_h}c^\dagger_{\Gamma,l,\mathbf{r}}T_{\mathbf{R}_h}^{-1} = (-1)^{l} c^\dagger_{\Gamma,l,\mathbf{r+R_h}}.
\end{equation}
The layer-dependent sign $(-1)^{l}$ arises because the fractional translation $T_{\mathbf{R}_h}$ shifts the local stacking configuration from sublattice $a$ of the top layer aligned with sublattice $a$ of the bottom layer, to sublattice $a$ of the top layer aligned with sublattice $b$ of the bottom layer. The monolayer band edge is composed of the antibonding combination of sublattices $a$ and $b$. This symmetry imposes a strict selection rule: $V_{\Gamma,l}(\mathbf{r})$ expands only in ``even'' moir\'e reciprocal lattice vectors $\mathbf{G}' \in \{ m\mathbf{g}'_{1} + n\mathbf{g}'_{2} \mid m,n \in \mathbb{Z},\ m+n\ \text{is even} \}$, while $\Delta_{\Gamma,T}(\mathbf{r})$ expands in ``odd'' vectors $\mathbf{G}' \in \{ m\mathbf{g}'_{1} + n\mathbf{g}'_{2} \mid m,n \in \mathbb{Z},\ m+n\ \text{is odd} \}$,
\begin{equation}\label{EMeq7}
    \begin{aligned}
        V_{\Gamma,l}(\mathbf{r}) =& \sum_{\mathbf{G}'\in \text{even}}e^{i\mathbf{G}'\cdot\mathbf{r}}V_{\Gamma,l}(\mathbf{G}'),\\
        \Delta_{\Gamma,T}(\mathbf{r})=&\sum_{\mathbf{G}'\in \text{odd}}e^{i\mathbf{G}'\cdot\mathbf{r}}\Delta_{\Gamma}( \mathbf{G'}).
    \end{aligned}
\end{equation}
Crucially, this constraint forces the zeroth-order (constant) tunneling term to vanish in the $\Gamma$-valley representation.

\subsubsection{Equivalence between M-valley and $\Gamma$-valley description and unified tunability}
A direct comparison shows that the Fourier expansions for $V_{\text{M},l}$ and $V_{\Gamma,l}$ (and similarly for $\Delta_{\text{M},T}$ and $\Delta_{\Gamma,T}$) contain exactly the same terms. Moreover, the rotational and time-reversal symmetries impose identical constraints on both expansion coefficients. Mathematically, the folded M-valley Hamiltonian is equivalent to the $\Gamma$-valley Hamiltonian:
\begin{equation}\label{EMeq8}
    \mathcal{H}^{\text{M},\text{folded}}_{0} = \begin{pmatrix}
        \mathcal{H}^\text{M}_0(\mathbf{k}) & \\
        & \mathcal{H}^\text{M}_0(\mathbf{k}+\mathbf{q}_{\text{M}})
    \end{pmatrix}\cong\mathcal{H}^\Gamma_0(\mathbf{k}),
\end{equation}
provided the expansion coefficients in the two descriptions are set equal. Thus, the M-valley system is a high-symmetry realization of the $\Gamma$-valley formalism subject to additional symmetry-enforced constraints.

We analyze the origin of Hubbard model tunability within this unified framework. By neglecting intralayer potentials $V_{\Gamma,l}$ and $V_{\text{M},l}$, the system possesses an emergent layer-exchange symmetry, thus the layer-bonding and antibonding states are eigenstates.
\begin{itemize}
    \item $\Gamma$-valley systems: A finite zeroth-order interlayer tunneling $\Delta_{\Gamma, T}(\mathbf{r})$ splits the bonding and antibonding states. An external displacement field hybridizes these states, leading to a widely tunable hopping ratio $t'/t$ well-described by non-degenerate perturbation theory.
    \item M-valley systems: The vanishing zeroth-order tunneling results in exact degeneracy between bonding and antibonding states. The displacement field induces coupling between these states and induces corrections of the topmost degenerate bands, which must be treated via degenerate perturbation theory.
\end{itemize}
Because in the $\Gamma$-valley description of M-valley systems, the unit-cell doubling is artificial, these two degenerate bands cannot be fully gapped by displacement field and must be unfolded into a single band in the original mBZ, which corresponds to the topmost single band in Ref.~\cite{Eugenio2025} and exhibit wide tunability. 

We conclude that in both systems, the tunability stems from the displacement-field-induced breaking of layer-exchange symmetry and the resulting considerable hybridization of layer-bonding and antibonding states.

\end{document}